\def\year{2021}\relax
\documentclass[letterpaper]{article} 

 \usepackage{subfigure}

\usepackage{aaai21}  
\usepackage{times}  
\usepackage{helvet} 
\usepackage{courier}  
\usepackage[hyphens]{url}  
\usepackage{graphicx} 
\def\UrlFont{\rm}  
\usepackage{natbib}  
\usepackage{caption} 
\newcommand{\Figref}[1]{Figure~\ref{#1}}

\title{Tactical Reframing of Online Disinformation Campaigns\\Against The Istanbul Convention}
\author {
    Tuğrulcan Elmas,
    Rebekah Overdorf,
    Karl Aberer\\
}
\affiliations {
	EPFL \\
    tugrulcan.elmas@epfl.ch, rebekah.overdorf@epfl.ch, karl.aberer@epfl.ch
}

\begin{document}

\maketitle

\begin{abstract}

In March 2021, Turkey withdrew from The Istanbul Convention, a human-rights treaty that addresses violence against women, citing issues with the convention's implicit recognition of sexual and gender minorities. In this work, we trace disinformation campaigns related to the Istanbul Convention and its associated Turkish law that circulate on divorced men's rights Facebook groups. We find that these groups adjusted the narrative and focus of the campaigns to appeal to a larger audience, which we refer to as ``tactical reframing.''  Initially, the men organized in a grass-roots manner to campaign against the Turkish law that was passed to codify the convention, focusing on one-sided custody of children and indefinite alimony. Later, they reframed their campaign and began attacking the Istanbul Convention, highlighting its acknowledgment of homosexuality. This case study highlights how disinformation campaigns can be used to weaponize homophobia in order to limit the rights of women. To the best of our knowledge, this is the first case study that analyzes a narrative reframing in the context of a disinformation campaign on social media. 

\end{abstract}

\label{sec:related}
\section{Background and Introduction}

First ratified by Turkey in 2011, \emph{The Council of Europe Convention on Preventing and Combating Violence Against Women and Domestic Violence}, better known as the \emph{Istanbul Convention}, has since been ratified by 34 European states in an effort to protect women from domestic violence. To ensure that the treaty was honoured, Turkey codified the convention via the \emph{Law to Protect Family and Prevent Violence against Woman}, colloquially referred to as \emph{Law No. 6284}. The convention and the law provide measures that prevent violence against women, protect victims of domestic abuse, and prosecute the perpetrators~\cite{conventionfactsheet}. 

In 2020, Numan Kurtulmuş, a representative of the Turkish ruling party, \emph{AKP}, stated that, due to demand from the public, they were considering withdrawing from the convention~\cite{duvarkurtulmus}. However, contemporary surveys found that only 7-9\% of participants supported the withdrawal from the convention, while 52\% of participants did not know what the convention was about~\cite{euronews,konda}. These remarks sparked a public debate. Opposition supporters and even from some pro-government organizations~\cite{duvarkadem} denounced the comments and began the campaign "İstanbul Sözleşmesi Yaşatır" (The Istanbul Convention Saves Lives)~\cite{duvarcampaign}. On the other side, pro-AKP and pro-SP (Islamist opposition party) political groups, religious groups (cults) and Islamist media argued that the convention was imposed by the west, normalized homosexuality~\cite{bianet}, increased divorces and hence destroys families, and increased the number of femicides~\cite{dogruhaber}.

On 20 March 2021, Turkey announced its withdrawal from the convention. The government announced that ``the convention was hijacked by a group of people attempting to normalize homosexuality, which is incompatible with Turkey's social and family values\footnote{https://www.iletisim.gov.tr/english/haberler/detay/statement-regarding-turkeys-withdrawal-from-the-istanbul-convention}.''

In this paper, we focus on anti-women's rights disinformation campaigns pushing for withdrawal from the Istanbul Convention. We found that these campaigns were organized by and for divorced men in Facebook groups. We show how the organizers reframed their arguments around homosexuality and the Istanbul Convention to appeal to a broader audience and garner support from different groups. 

Using Facebook data provided by CrowdTangle, we show how these groups organized in a grass-roots manner to campaign against Law No. 6284, which they claimed overprotected their ex-wives by allowing one-sided custody of children to women in divorce cases and by allowing for indefinite alimony. They later reframed their campaign around attacking the Istanbul Convention, citing that it promotes homosexuality. This reframing allowed for a broader audience, including religious and conservative groups. 

\section{Dataset Methodology}
\label{sec:collection}
\begin{figure*}[ht]
    \centering
    \includegraphics[width=\linewidth]{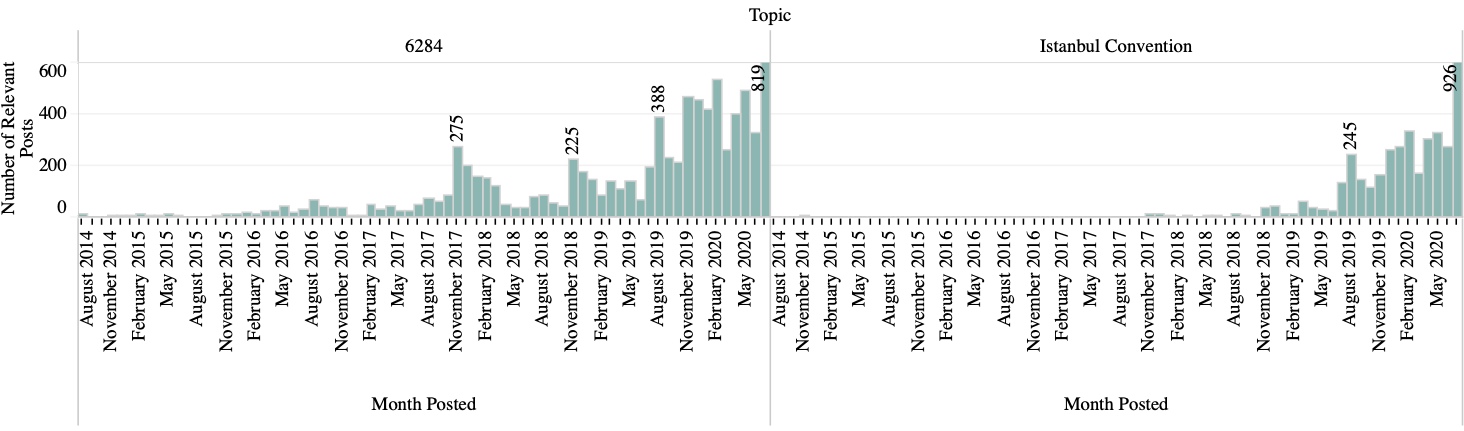}
    \caption{Time Series of posts containing ``6284'' (left) and ``Istanbul Convention'' (in Turkish, right). Posts are grouped by the month in which they were posted (``Created At'').}
    \label{fig:fbdata}
\end{figure*}
Our analysis leverages data from Facebook, provided by CrowdTangle~\cite{crowdtangle}. CrowdTangle allows academics and researchers to collect public posts from public pages, groups and verified profiles on Facebook, prioritizing those who work on misinformation, elections, COVID-19, racial justice and well-being\footnote{https://help.crowdtangle.com/en/articles/4302208-crowdtangle-for-academics-and-researchers}. We focus on Facebook because Facebook posts are intrinsically labelled according to their source, i.e., since we know where a post originated, we know the post's motivation via the motivation of the page/group. Additionally, unlike Twitter where some users post to create popularity and/or reach non-followers, Facebook is better suited for communicating with friends and among members of groups or followers of pages, making it more appropriate for analyzing how users communicate when organizing. 

We aimed to capture the grass-roots aspect of this campaign to better understand the organizing that occurred on social media. The moment that Kurtulmuş announced that the ruling party was considering withdrawing from the convention (July 2020) the campaign became a top-down effort, so we collected and analyzed data prior to that moment. The announcement also created a backlash, causing an outpouring of pro-convention content into the post-announcement conversation. 

We used two queries to find relevant data: ``İstanbul Sözleşmesi'' (\emph{Istanbul Convention}) and ``6284'' (a reference to Law No. 6284, which ensures the Istanbul Convention is honoured). For CrowdTangle to return data in a search, a researcher using the service must have tracked the group/page/profile that contains the data. To this end, we manually curated a list of groups/pages/profiles that either contain at least one post containing ``İstanbul Sözleşmesi'' or ``6284'' or are related to Turkish politics and media alongside an anti-Istanbul convention stance (e.g. groups related to political parties, media and religious cults). This resulted in 2,500 groups/pages/profiles. We collected all public posts that CrowdTangle offered containing the keywords, discarding non-Turkish posts. We found that noise introduced by using a number as a query was negligible: Turkish posts mentioning 6284 are almost exclusively about Law No. 6284, so we abstained from introducing additional filters that may have harmed recall.

\section{Analysis}

\begin{figure}[ht]
\centering     
\subfigure{\label{fig:a}\includegraphics[width=\columnwidth]{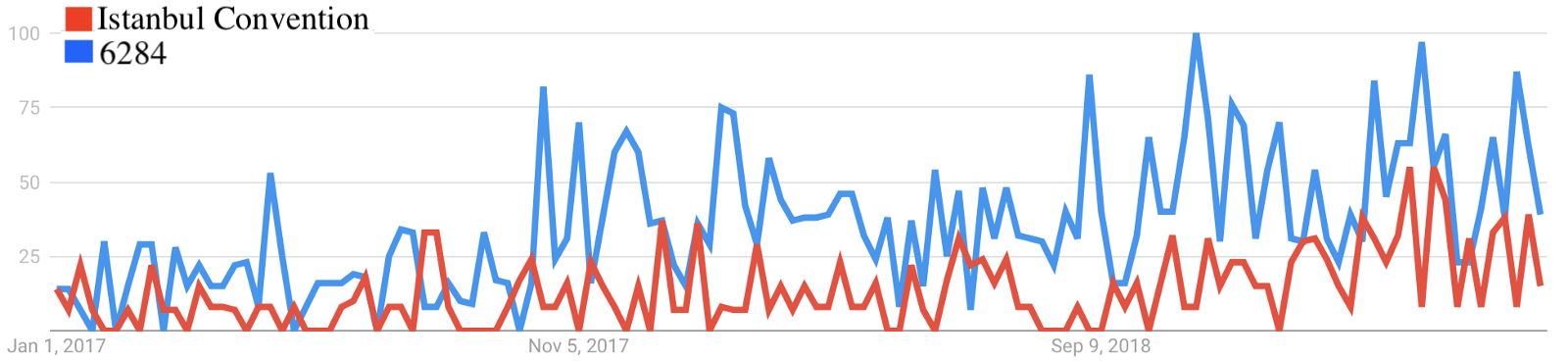}} \\	
\subfigure{\label{fig:c}\includegraphics[width=\columnwidth]{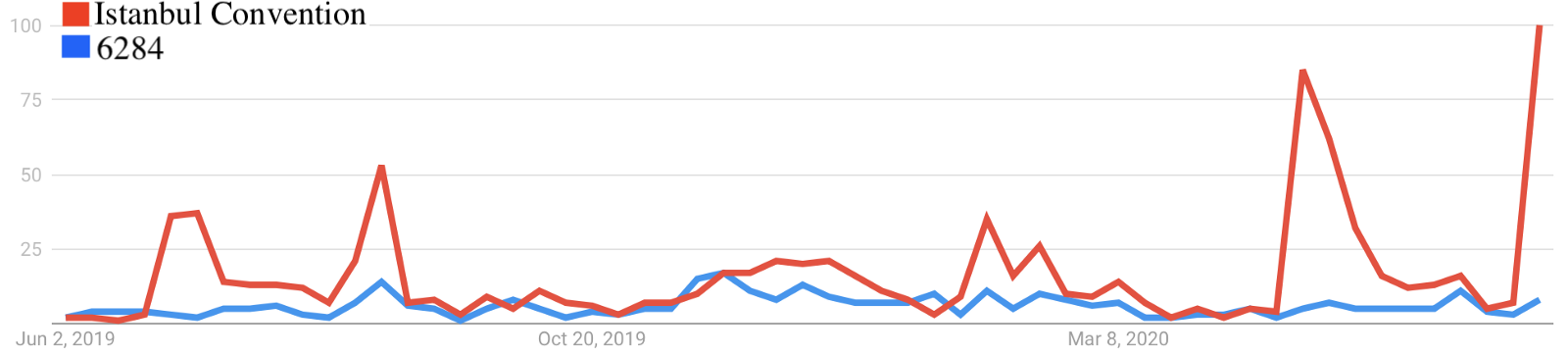}}

\caption{Google trends data. Between 2017-2019 Law No. 6284 was searched more than the convention. This shifted towards the Istanbul Convention in 2019. This aligns with our findings from Facebook.
\label{fig:googletrends}}        
\end{figure}

In our analysis, we observe a temporal change in the topics discussed in the divorced men's groups manifesting a tactical reframing. We first discuss the summary statistics of the collected data. \Figref{fig:fbdata} shows the number of relevant posts per month with respect to the two topics (6284 and The Istanbul Convention). We observe that there were few but consistent posts mentioning 6284 between 2015-2018, with two peaks in November 2017 and November 2018, signalling that there was some discussion related to the law. Meanwhile, these peaks are absent from the Istanbul Convention discussion; the convention was not consistently discussed until the latter half of 2019, with its first peak in August 2019 coinciding with a peak in the discussion about Law No. 6284. It is clear that the debate related to Law No. 6284 precedes the debates related to the Istanbul Convention. We complement this finding with Google trends data, to show that it's not a platform-specific feature nor an artefact of the dataset. \Figref{fig:googletrends} shows that more public attention was initially focused on 6284 rather than the Istanbul Convention, but that this shifts in 2019. 

\pagebreak

While the analysis shows a growing interest and attention to the topic over time, it remains unclear whether this discussion originated on and was fueled by Facebook itself or whether it was instead the consequence of public interest induced by a general discussion on media, newspapers, etc. That is, is the Facebook discussion only a byproduct of the broader discussion or did it have a hand in directing the conversation? We leave further analysis of this to future work, e.g. to compare the time series of the two keywords as obtained from Facebook posts and from Google Trends data to determine whether there is a correlation that confirms the public interest in the topic.

\subsection{Source of Posts}

Numan Kurtulmuş cited a ``public demand'' to withdraw from the convention, suggesting that there was bottom-up organizing against the convention. We found that there was bottom-up organizing against the convention prior to his statement, however, the motivations of those pages and groups differed from the official motivation of the government's withdrawal from the treaty.

\begin{figure}[b!]
    \centering
    \includegraphics[width=\columnwidth]{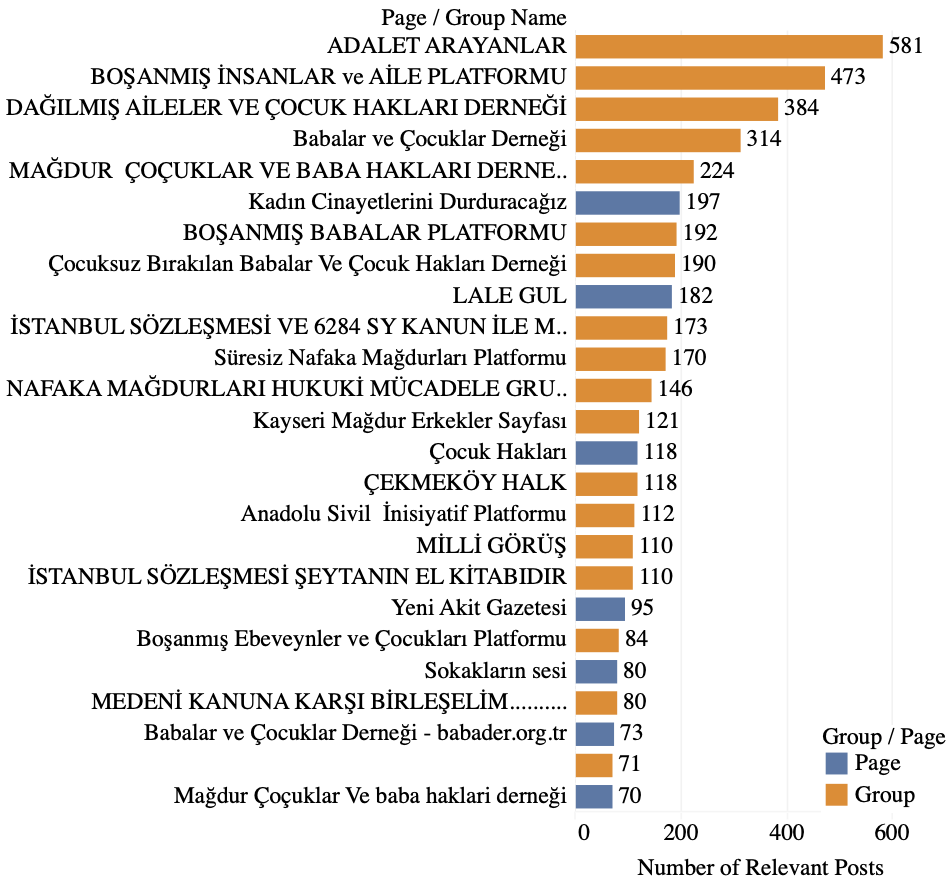}
    \caption{The most active groups and pages with at least 70 posts related to Law No. 6284 or The Istanbul Convention. The top 5 most active groups are (translated): ``Those who seek justice,'' ``The platform for family and divorced people,'' ``Broken homes and children rights association,'' ``Fathers and children association,'' and ``Victim kids and father rights association.'' Some groups were not oriented to divorced men:``Milli Görüş'' and ``Milli Görüş Erleri'' are political groups, ``LALE GUL'' (Tulip and Rose) is a cult-related group posting Islamic content, ``Çekmeköy Halk'' and ``Anadolu Sivil İnsiyatifi Platformu'' are location-based groups without a primary topic.}
    \label{fig:mostactive}
\end{figure}

\begin{figure}[th]
    \centering
    \includegraphics[width=\columnwidth]{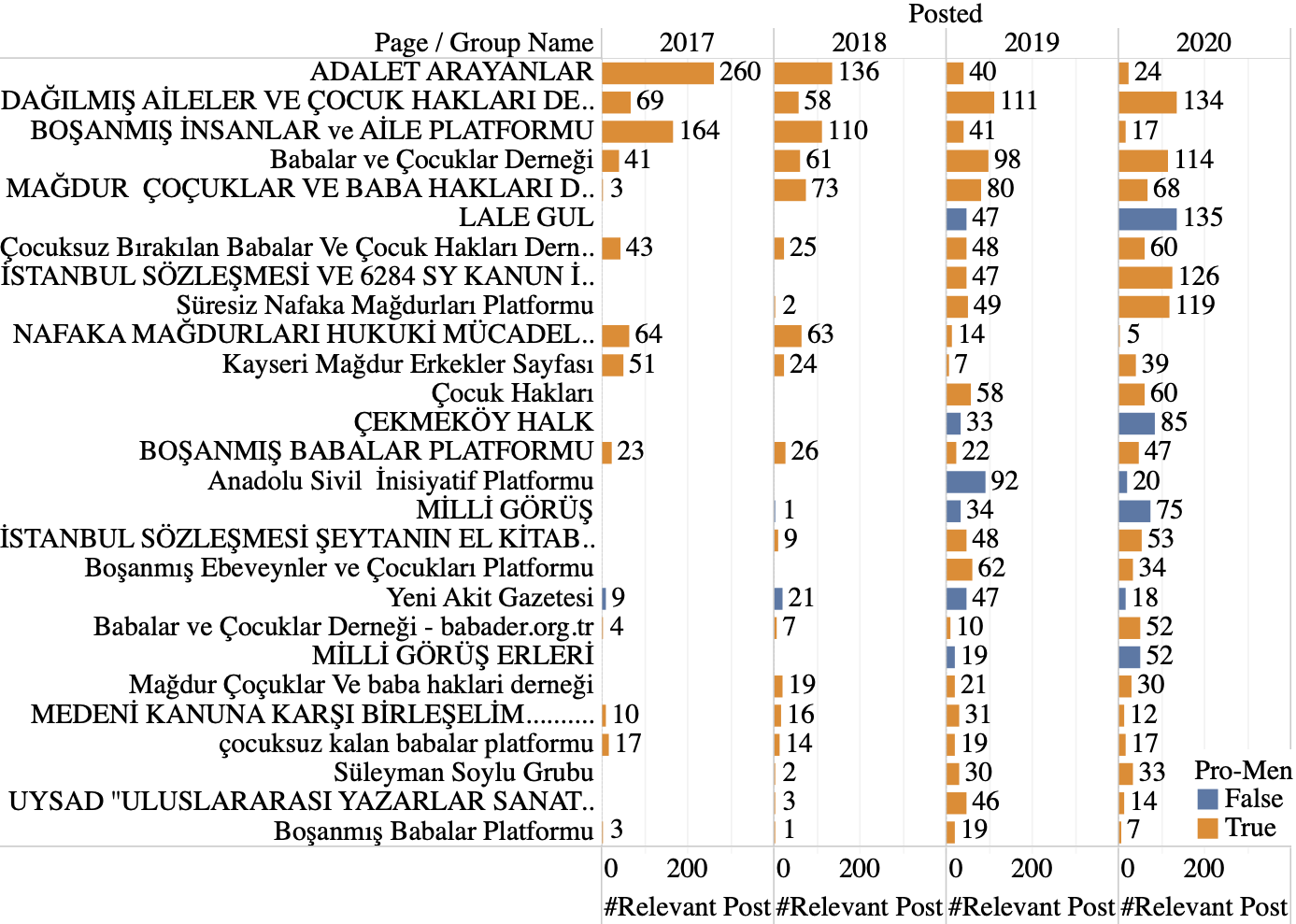}
    \caption{While the groups/pages of divorced men were active before 2019, all the other groups (shown in blue) joined the conversation later in 2019 except for the Yeni Akit Gazetesi, an Islamist newspaper.}
    \label{fig:mostactiveyears}
\end{figure}

We found that most of the groups actively discussing the Istanbul Convention or Law No. 6284 self-stated that they were concerned with the rights of divorced men. We classify these groups as "pro-men" by inspecting the name of the page or the group and/or the about section of the page or the group, as all these groups self-state their motivations in these two sections while other groups cite other motivations in these fields. \Figref{fig:mostactive} shows the most active groups in the discussion, sorted by the number of relevant posts, which mention either the Istanbul Convention or Law No. 6284. ~\Figref{fig:mostactiveyears} shows these groups' activities per year. With one exception, the divorced men's groups are the one's discussing the Istanbul Convention/Law No. 6284 before 2019, with other groups joining the conversation in late-2019. 

\subsection{Tactical Reframing of The Campaign}
\label{sec:exploratory}

\begin{figure}[ht]
    \centering
    \includegraphics[width=\columnwidth]{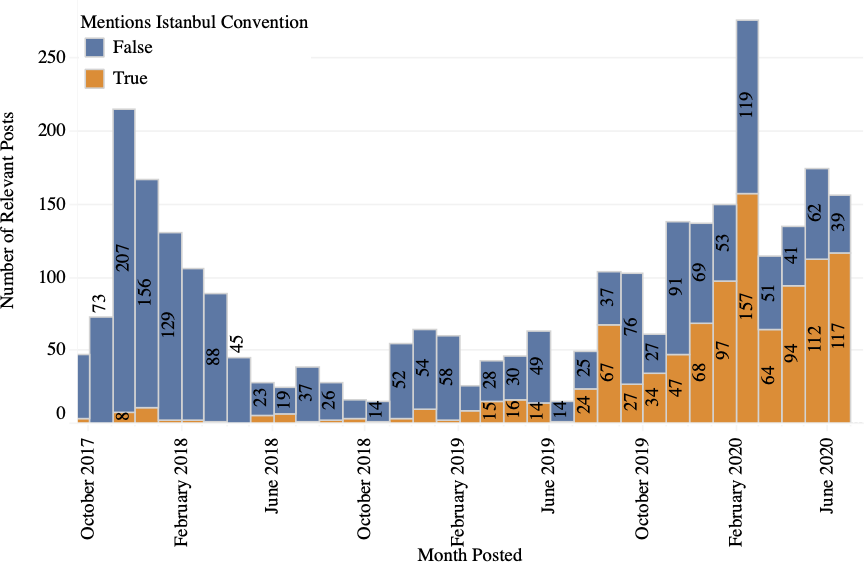}
    \caption{The time series of posts mentioning the Istanbul Convention or ``6284'' from 35 groups and pages related to divorced men's rights (manually annotated by examining  their name and about page). The \texttt{False} class represents posts that only reference 6284, and the \texttt{True} class shows any mention of the Istanbul Convention.}
    \label{fig:shift1}
\end{figure}

Not only do more groups begin discussing the Istanbul Convention/Law No. 6284 in 2019, but mentions of the Istanbul Convention begin to overtake discussions of Law No. 6284. This is partially due to political and religious groups joining the discussion, however, as shown in \Figref{fig:shift1}, the divorced men's groups also shift to target the Istanbul Convention (both on its own and together with Law No. 6284). By late 2019 the number of posts about The Istanbul Convention surpassed those about Law No. 6284. 

We also see this tactical reframing in how some of the groups' change names. For instance, the group ``ISTANBUL SÖZLEŞMESİ ŞEYTANIN EL KİTABIDIR'' (\emph{The Istanbul Convention is the Book of the Devil}) appears to have formed against the convention for religious reasons, however, initially, the group was named ``6284 KALKSIN'' (\emph{Abolish 6284}) and its first post reads (translated): 
\begin{quote}
     \emph{Dear friends, this group was established to defend the rights of parents who could not reach their children due to legal gaps. We kindly ask you to add other victims like ourselves (to the group)}.
\end{quote}

\noindent Similarly, another group initially created in 2018 under the name ``boşanmış insanlar platformu'' (\emph{the platform of divorced people}) later changed its name to ``İSTANBUL SÖZLEŞMESİ VE 6284 SY KANUN İLE MÜCADELE PLATFORMU'' (\emph{The Platform to Fight against the Istanbul Convention and Law No. 6284}).

\begin{figure}[htb!]
    \centering
    \includegraphics[width=\columnwidth]{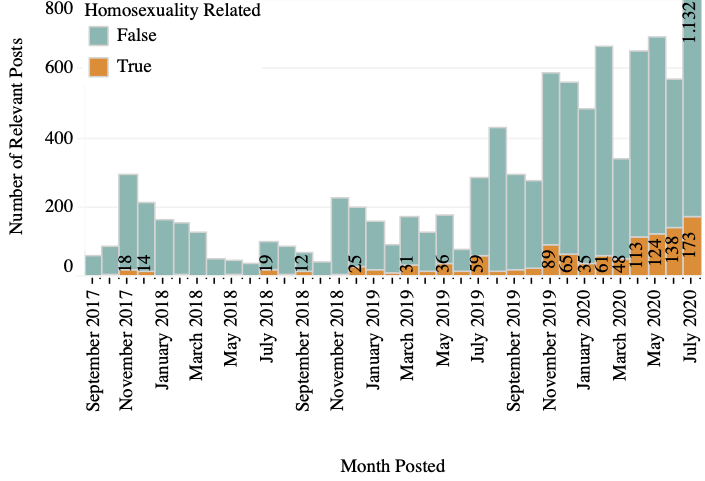}
    \caption{The number of posts containing homosexuality related substrings: ``homo'', ``escinsel'' and ``LGBT'' (in orange) among all posts discussing The Istanbul Convention or Law No 6284. Homosexuality related posts are rare until 2019, when the convention becomes the main target.}
    \label{fig:homosexuality_debate_shift}
\end{figure}

While of course it is never explicitly stated why this shift occurred, Law No. 6284 was established as a result of the Istanbul Convention and withdrawing from the latter may open the door to the repeal of the former. Additionally, the Istanbul Convention is an easier target than Law No 6284: it comes from the west, a regular target of Islamists, and it contains articles explicitly mentioning sexual orientation, equality of women and men, and the rejection of conservative ideas of gender roles, the fear of which are exploited. For example, convention protesters leverage homophobia by pointing to article 4/3 which mentions ``sexual orientation'' and thus implicitly acknowledges homosexuality, and they leverage fear over the breakdown of conservative gender roles by claiming that article 14/1, which mentions non-stereotyped gender roles, implicitly rejects the roles of men and women according to Islam. Indeed, posts mentioning homosexuality (which was the official reason given by the government to withdraw) increased alongside mentions of the Istanbul Convention, though they still made up only a small part of the debate, as shown in ~\Figref{fig:homosexuality_debate_shift}.

\subsection{Evidence of Interplay}

In the previous section, we demonstrated that there was a shift in the focus of the divorced men group's frames from Law No. 6284 to The Istanbul Convention which coincided with an increase in posts mentioning homosexuality. We also observed that the increase in the number and variety of groups discussing the Istanbul Convention coincides with this shift. Essentially, we have demonstrated a correlation between these events. Next, we explore the interplay between these phenomena. We hypothesize that the shift was an attempt to appeal to more conservative, Islamist, and homophobic populations and provide evidence to this effect. In short, we find that the divorced men's groups collaborated with Yeni Akit, an Islamist pro-government newspaper that was reported to be campaigning against the convention~\cite{bianet}. 

\begin{figure}[b!]
    \centering
    \includegraphics[width=\columnwidth]{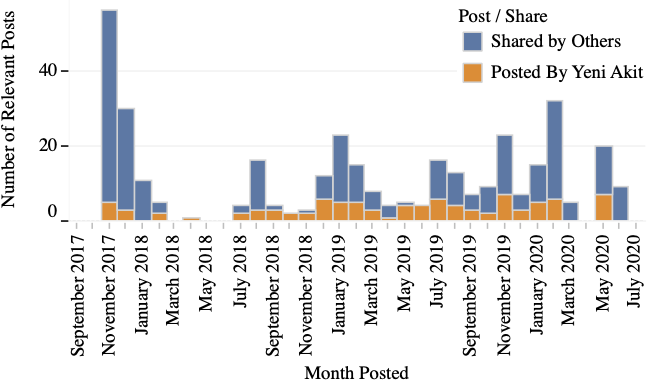}
    \caption{The number of Yeni Akit news articles that mention ``6284'' or the Istanbul Convention shared by the Yeni Akit's Facebook group (orange) and by other groups (blue).}
    \label{fig:yeniakitposts}
\end{figure}

First, we see in \Figref{fig:yeniakitposts} that the divorced men's groups consistently share Yeni Akit news stories about Law No. 6284 and The Istanbul Convention. In several cases, users in divorced men's groups shared the news while citing the groups' success in campaigning and thanking the journalist for helping their cause, implying that the articles were the result of their efforts. For example, a news story from 2018 stated that people were fed up with Law No. 6284 so they were campaigning against it on social media by describing their problems and that Yeni Akit was helping them in setting the agenda~\cite{yeniakitagenda}. One of the groups shared this article and thanked the author, who replied that it was his pleasure to support their cause and claimed that the law was imposed by the west to destroy families in Turkey.

In a similar story, Yeni Akit promoted anti-Law No. 6284 men campaigning against indefinite alimony, stating that ``people shook social media with the hashtag \#SüresizNafakaZulümdür (\emph{Alimony for an Indefinite Period is Torture}) which became a top trend (on Twitter).''~\cite{yeniakitalimony}. Prior work has shown that the hashtag did indeed reach trending, but that it was the result of bot activity~\cite{elmas2020power}.





\begin{figure}[h!]
    \centering
    \includegraphics[width=\columnwidth]{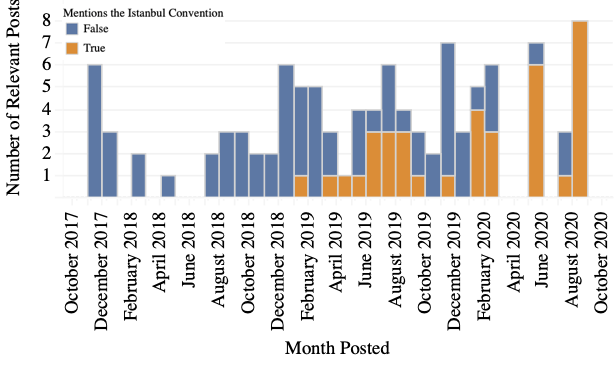}
    \caption{The Number of posts by the Yeni Akit Facebook group mentioning the Istanbul Convention either within the Facebook post text or the text in the article preview, which consists of the headline.}
    \label{fig:yeniakitreframe}
\end{figure}

We also witnessed the same shift from Law No. 6284 to the Istanbul convention in Yeni Akit, as ~\Figref{fig:yeniakitreframe} demonstrates. Here the shift is rather explicit. In one news story, they report that \emph{The Association of Fathers and Children} (one of the Facebook groups formed initially against Law No. 6284) stated that ``The Istanbul Convention is the keystone of Law No. 6284'', promoting the group and explicitly demonstrating the shift in target~\cite{yeniakitreframe}. In another article, they reported that an advocate campaigning against indefinite alimony stated that Turkey should withdraw from the Istanbul Convention and then change Law No. 6284 to save families~\cite{yeniakitadvocate}.



\subsection{The Impact of The Groups}

\begin{figure}[h!]
    \centering
    \includegraphics[width=\columnwidth]{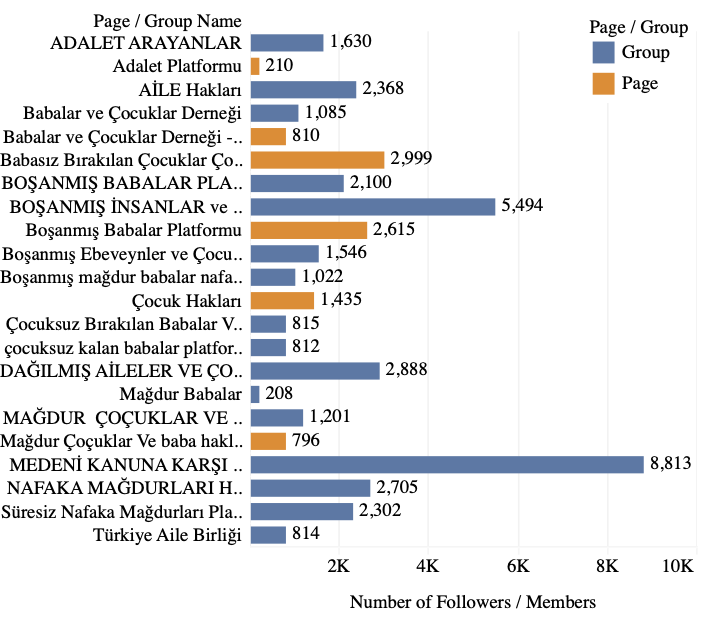}
    \caption{Number of followers/members per group/page related to divorced men's rights.}
    \label{fig:page_group_members}
\end{figure}

\begin{figure}[h!]
    \centering
    \includegraphics[width=\columnwidth]{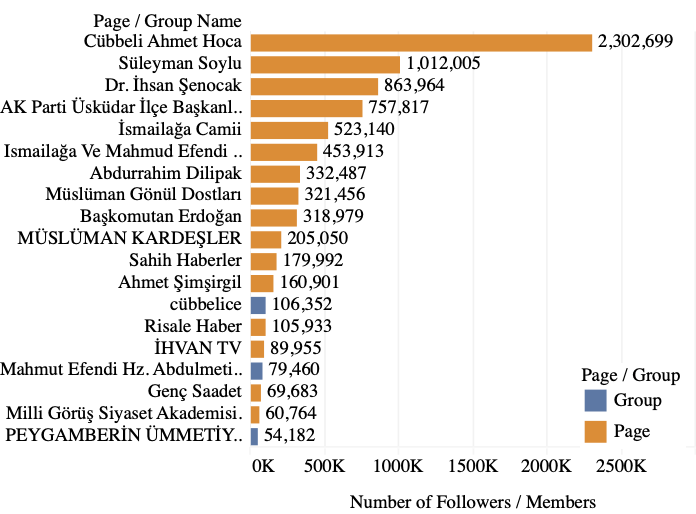}
    \caption{Number of followers/members per group/page which contain posts about Law No. 6284 or the Istanbul Convention. We exclude newspapers from this list as they generally only report about the convention. The rest of the pages are mostly related to politics and cults. The top 5 pages are: Cübbeli Ahmet Hoca (cult leader), Süleyman Soylu (AKP politician), Dr. İhsan Şenocak (cult leader), AK Party Üsküdar District Center, İsmailağa Camii (a mosque).}
    \label{fig:page_group_benchmark}
\end{figure}

We found that the pages and groups formed by divorced men actually have relatively few followers and members, usually between 800 and 3,000 as \Figref{fig:page_group_members} shows. While this implies that the groups likely have little influence on the decision to withdraw, it supports their motivation to undergo a tactical reframing: while the problems of divorced men only get attention from other divorced men, homophobia is common to many Islamists. Indeed, we found that the groups/pages with the most members/followers were all related to politics and religious cults as shown in \Figref{fig:page_group_benchmark}. These groups might have been effective at reaching a broader and more diverse audience in the anti-convention campaign.

Note that this is under the assumption that a page's/group's follower/member count is the only metric for impact. However, we consider Facebook data as only a window in their grassroots organizing and the tactical reframing of their campaign. Group members may also use closed channels like Whatsapp groups to communicate. Moving offline, some of the groups are also associations founded to lobby against Law No. 6284 and The Istanbul Convention, such as Babalar ve Çocuklar Derneği (\emph{Association of Fathers and Children})\footnote{http://babader.org}, Babasız Bırakılan Çocuklar Çocuksuz Bırakılan Babalar Derneği (\emph{Assocation of Fatherless Children and Childrenless Fathers}), Dağılmış Aileler ve Çocuk Hakları Derneği (\emph{Destroyed Families and Children Rights Association}). The men involved in these groups have also reportedly been involved in street protests~\cite{yeniakitprotest}. Therefore, it appears that these men actively lobby and might have more impact than their Facebook groups/pages suggest.

\section{Discussion}
\label{sec:usecases}

\subsection{Caveats}

Our analysis shows that the divorced men's groups were demanding the withdraw from The Istanbul Convention and that they weaponized homophobia and Islamists beliefs, however, the causation is not apparent. Our analysis remains inconclusive as to whether or not the divorced men's groups directly caused Islamists to campaign against the convention, although evidence supports this idea. Additionally, there is always a possibility that Islamists plotted against the convention as well through closed channels (e.g. Whatsapp groups), although the Facebook data reveals no such bottom-up campaign. 

\subsection{Future Work}

Our initial observations show that the campaign against the convention was initially focused on men's rights, but was reoriented to political and religious arguments and, reframed in a way that appealed to the public's emotions and fears, especially of homosexuality and non-traditional gender roles. In future work, we will explore this transition more deeply. 

In this analysis, we only reported aggregate statistics while showing that the debate has shifted towards the Istanbul Convention and homosexuality while the participating groups stayed the same. Another important question is, while the groups stayed the same, did the users stay the same as well? i.e. did the users reframe individually? We plan to address this question in future work through a fine-grained analysis of user behaviours within the groups and through employing data from other social networks. 
\subsection{Data Privacy Considerations}
\label{sec:ethics}

This study only utilises public groups and pages provided by CrowdTangle. We did not analyze nor store any other data, such as personal files.

\section{Related Work}

Disinformation campaigns on social media employ accounts with coordinated behaviour \cite{keller2020political,elmas2020power}. Past studies reveal that accounts used in disinformation campaigns might observe a shifting behaviour \cite{llewellyn2019whom,elmas2020misleading}. However, those studies reported on accounts with shifted identities and focus. This study is the first to report groups of accounts with the same identity and focus but employing different frames in a disinformation campaign. 

Frames highlight certain issues, events or beliefs versus others~\cite{benford2000framing}. Studies focusing on frames reported shifting frames in presidential speeches~\cite{arthur2013contextual} and news media~\cite{woods2014threat,greussing2017shifting}. However, to the best of our knowledge, no previous work analyzed shifting frames in the context of disinformation campaigns organized by social media users.

\fontsize{9.0pt}{10.0pt} \selectfont
\bibliography{bib}

\end{document}